\documentclass[11pt,
twocolumn,
 aip,
 apl,%
 showkeys,
 amsmath,amssymb,
reprint
]{revtex4-1}
\usepackage{sidecap}
\usepackage[breaklinks=true,colorlinks=true,linkcolor=blue,urlcolor=blue,citecolor=blue]{hyperref}
\usepackage{graphicx}%
\usepackage{graphics}%
\usepackage{epstopdf}
\usepackage{dcolumn}%
\usepackage{soul,color}
\usepackage{bm}
\usepackage[mathlines]{lineno}
\expandafter\ifx\csname package@font\endcsname\relax\else
 \expandafter\expandafter
 \expandafter\usepackage
 \expandafter\expandafter
 \expandafter{\csname package@font\endcsname}%
\fi
\usepackage[utf8]{inputenc}
\usepackage[T1]{fontenc}
\usepackage{mathptmx} 

\begin{document}
\abovedisplayskip=3pt
\belowdisplayskip=3pt
\abovedisplayshortskip=2pt
\belowdisplayshortskip=2pt


\title{\Large{Spin-wave eigenmodes in direct-write 3D nanovolcanoes}}
\author{O.~V. Dobrovolskiy}
    \email{oleksandr.dobrovolskiy@univie.ac.at}
    \affiliation{Faculty of Physics, University of Vienna, 1090 Vienna, Austria}
\author{N.~R.~Vovk}
    \affiliation{Institute of Physics for Advanced Materials, Nanotechnology and Photonics (IFIMUP)/Departamento de F\'isica e Astronomia, Universidade do Porto, 4169-007 Porto, Portugal}
    \affiliation{Department of Physics, V.\,N. Karazin Kharkiv National University, Svobody Sq. 4,
    Kharkiv 61022, Ukraine}
\author{A. V.~Bondarenko}
    \affiliation{Institute of Physics for Advanced Materials, Nanotechnology and Photonics (IFIMUP)/Departamento de F\'isica e Astronomia, Universidade do Porto, 4169-007 Porto, Portugal}
\author{S. A.~Bunyaev}
    \affiliation{Institute of Physics for Advanced Materials, Nanotechnology and Photonics (IFIMUP)/Departamento de F\'isica e Astronomia, Universidade do Porto, 4169-007 Porto, Portugal}
\author{S. Lamb-Camarena}
    \affiliation{Faculty of Physics, University of Vienna, 1090 Vienna, Austria}
\author{N. Zenbaa}
    \affiliation{Faculty of Physics, University of Vienna, 1090 Vienna, Austria}
\author{R.~Sachser}
    \affiliation{Physikalisches Institut, Goethe University, 60438 Frankfurt am Main, Germany}
\author{S.~Barth}
    \affiliation{Physikalisches Institut, Goethe University, 60438 Frankfurt am Main, Germany}
\author{K.~Y.~Guslienko}
    \affiliation{Division de Fisica de Materiales,
      Depto. Polimeros y Materiales Avanzados: Fisica, Quimica y Tecnologia,
      Universidad del Pais Vasco, UPV/EHU,
      20018 San Sebastian, Spain}
    \affiliation{IKERBASQUE, the Basque Foundation for Science,
     48009 Bilbao, Spain}
\author{A.~V. Chumak}
    \affiliation{Faculty of Physics, University of Vienna, 1090 Vienna, Austria}
\author{M. Huth}
    \affiliation{Physikalisches Institut, Goethe University, 60438 Frankfurt am Main, Germany}
\author{G.~N. Kakazei}
    \affiliation{Institute of Physics for Advanced Materials, Nanotechnology and Photonics (IFIMUP)/Departamento de F\'isica e Astronomia, Universidade do Porto, 4169-007 Porto, Portugal}
\date{\today}

\begin{abstract}
Extending nanostructures into the third dimension has become a major research avenue in modern magnetism, superconductivity and spintronics, because of geometry-, curvature- and topology-induced phenomena. Here, we introduce Co-Fe nanovolcanoes---nanodisks overlaid by nanorings---as purpose-engineered 3D architectures for nanomagnonics, fabricated by focused electron beam induced deposition. We use both perpendicular spin-wave resonance measurements and micromagnetic simulations to demonstrate that the rings encircling the volcano craters harbor the highest-frequency eigenmodes, while the lower-frequency eigenmodes are concentrated within the volcano crater, due to the non-uniformity of the internal magnetic field. By varying the crater diameter, we demonstrate the deliberate tuning of higher-frequency eigenmodes without affecting the lowest-frequency mode. Thereby, the extension of 2D nanodisks into the third dimension allows one to engineer their lowest eigenfrequency by using 3D nanovolcanoes with 30\% smaller footprints. The presented nanovolcanoes can be viewed as multi-mode microwave resonators and 3D building blocks for nanomagnonics.
\end{abstract}
\maketitle

Extending nanomagnets into the third dimension has become a vibrant research avenue in modern magnetism \cite{Fer17nac,Fer20mat}. It encompasses investigations of 3D frustrated systems \cite{Kel18nsr,May19cph,Gli20apl,Skj20nrp}, topology- and curvature-induced effects in complex-shaped nano-architectures \cite{Str16jpd,Vol19prl,San20acs,She20cph}, and the dynamics of spin waves in 3D magnonic systems \cite{Kra08prb,Yan11apl,Ota16prl,Gub19boo,Sak20apl}. In magnonics, which is concerned with the operations with data carried by spin waves, magnonic conduits have traditionally been made from 2D structures \cite{Kru10jpd,Dem13boo,Liu13apl,Kak14apl,Wan20nel,Liy20jap}. Extension of spin-wave circuits into the third dimension is required for the reduction of footprints of magnonic logic gates \cite{Wan20nel} and it allows, e.g., steering of spin-wave beams in graded-index magnonics \cite{Dav15prb,Toe16nsr,Gra17prb}. In addition, the height of nanomagnets offers an additional degree of freedom in the rapidly developing domain of inverse-design magnonics \cite{Wan20arh,Pap20arx} in which a device functionality is first specified, then a feedback-based computational algorithm is used to obtain the device design. In the past, peculiarities of the lithographic process were used, e.g., for the formation of crowns on the tops of nanodisks \cite{Ste11ras}. However, lithographic techniques insufficiently suit the demands of 3D magnonics. This motivates the increasingly growing attention to additive manufacturing nanotechnologies \cite{Fer20mat}.

In recent years, two-photon 3D lithography and 3D direct writing by focused electron and ion beam-induced deposition (FEBID and FIBID) have become the techniques of choice for the fabrication of complex-shaped nano-architectures in magnetism, superconductivity and plasmonics \cite{Kel18nsr,May19cph,Fer20mat,Win17ami,Kel18nsr,Dob18nac,Cor19nal,Por19acs,Win19jap,Dob19pra,May19cph,Fer20mat,San20acs}. For magnonics, the propagation of spin waves in direct-write Fe- and Co-based conduits has recently been demonstrated, with a spin-wave decay length in the range $3$-$6$\,$\mu$m \cite{Dob19ami,Fla20prb}. Given FEBID's lateral resolution down to $10$\,nm and its versatility regarding the substrate material, FEBID appears as a promising nanofabrication technology for 3D magnonics \cite{Die20nel}.

Here, we investigate the spin-wave eigenmodes in individual direct-write Co-Fe nanovolcanoes by spin-wave resonance (SWR) spectroscopy \cite{Dob20nan,Bun21apl} and analyze the experimental data with the aid of micromagnetic simulations. We reveal that the microwave response of the nanovolcanoes essentially differs from the sum of the microwave responses of their constituent 2D elements---nanorings and nanodisks. We demonstrate that the ring encircling the volcano crater leads to an effective confinement of the low-frequency eigenmodes under the volcano crater, while the higher-frequency eigenmodes are confined in the ring area. By varying the crater diameter by $\pm20$\,nm, we demonstrate the deliberate tuning of the higher eigenfrequencies by about $\pm2$\,GHz without essential variation of the lowest eigenfrequency. The presented nanovolcanoes can be viewed as multi-mode resonators and as 3D building blocks for nanomagnonics.
\begin{figure}[t!]
    \centering
    \includegraphics[width=0.95\linewidth]{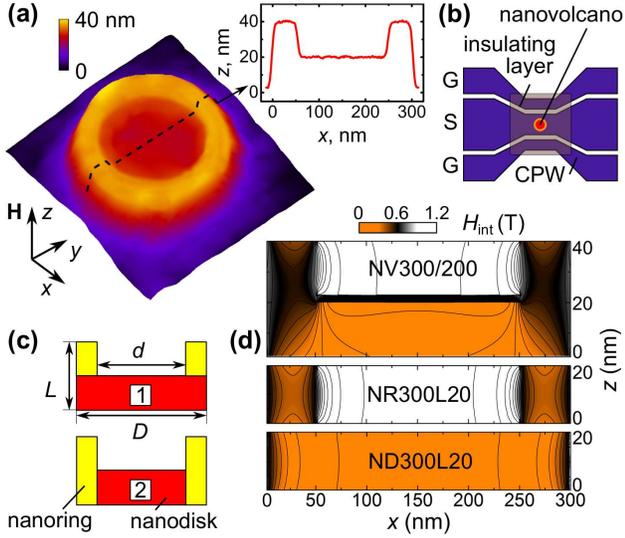}
    \caption{Experimental geometry.
    (a) Atomic force microscopy image of the nanovolcano NV$300/200$ with the outer diameter $D=300$\,nm and the crater diameter $d=200$\,nm. Inset: cross-sectional line scan, as indicated.
    (b) Active part of the coplanar waveguide (S: signal; G: ground) with a nanovolcano for microwave spectroscopy measurements (not to scale).
    (c) Possible ways of splitting a nanovolcano into a combination of a disk and a ring for modeling.
    (d) Calculated spatial dependence of the internal field $H_\mathrm{int}$ for the 40\,nm-thick nanovolcano NV300/200 and its individual basic elements (nanodisk ND300L20 and nanoring NR300L20).}
    \label{f1}
\end{figure}
\begin{table}[b!]
{\small
\begin{tabular}{|l|cccccccc|}
\hline
Sample\,$D/d$,\,& $L$,  &   $M_\mathrm{s}$, &   $A$,   &   D1,   &   D2,   &   D3,   &   R1,    &   R2,  \\
   nm           & nm    &  \,kA/m\,& \,pJ/m\, & \,GHz\, & \,GHz\, & \,GHz\, & \,GHz\,   & \,GHz\,\\
\hline
NV$1000/700$    & 40    &   1125   &   16.7   &   10.19 &   11.70 &   13.11 &   20.22   &   21.05\\
NV$600/400$     & 40    &   1030   &   16.1   &   14.57 &   17.04 &   19.04 &   26.93   &   28.58\\
NV$300/180$     & 40    &   880    &   15.4   &   7.29  &   12.06 &   17.85 &   22.01   &   23.66\\
NV$300/200$     & 40    &   880    &   15.4   &   7.32  &   11.67 &   17.01 &   21.95   &   25.16\\
NV$300/220$     & 40    &   880    &   15.4   &   7.38  &   11.37 &   16.25 &   25.95   &   27.71\\
\hline
NR$300/200$     & 20    &   880    &   15.4   &   --    &   --    &   --    &   --      &   17.72\\
NR$300/200$     & 40    &   880    &   15.4   &   --    &   --    &   --    &   --      &   24.89\\
ND$200$         & 20    &   880    &   15.4   &   10.47 &   15.83 &   21.68 &   --      &   -- \\
ND$300$         & 20    &   880    &   15.4   &   7.92  &   11.37 &   15.05 &   --      &   -- \\
ND$340$         & 20    &   880    &   15.4   &   7.32  &   10.44 &   13.65 &   --      &   -- \\
\hline
\end{tabular}
}
\caption{Sample parameters and simulated SWR frequencies for nano-volcanoes, nano-rings and nano-disks. NV: nanovolcano; ND: nanodisk; NR: nanoring; $D$: outer diameter; $d$: inner diameter; $L$: thickness; $M_\mathrm{s}$: saturation magnetization; $A$: exchange stiffness. $M_\mathrm{s}$ and $A$ values correspond to those for equivalent diameter Co-Fe nanodisks \cite{Dob20nan}. The frequencies of the modes D1-D3, R1 and R2 are calculated at $1.7$\,T for NV$1000/700$ and NV$600/400$, and at $1.2$\,T for all other samples. Upper part: Nanovolcanoes studied experimentally and by micromagnetic simulations. Lower part: Nanovolcanoes' individual elements (nanodisks and nanorings) studied by simulations.}
\label{t1}
\end{table}

The 40\,nm-thick Co-Fe nanovolcanoes with diameters down to $300$\,nm and crater diameters down to $180$\,nm were fabricated by FEBID (SEM: FEI Nova NanoLab 600) on top of a gold coplanar waveguide (CPW), see Fig.\,\ref{f1}. FEBID was done with $5$\,kV/$1.6$\,nA, $20$\,nm pitch, $1\,\mu$s dwell time, HCo$_3$Fe(CO)$_{12}$ as precursor gas \cite{Por15nan,Kum18jpc} and a serpentine scanning strategy \cite{Bun21apl}. Five nanovolcanoes with different outer and inner (crater) diameters, $D$ and $d$, respectively, were fabricated (see Table \ref{t1}). We label the nanovolcanoes with their diameter ratios NV$D/d$\,(nm) and use the prefixes ND and NR for the nanodisks and nanorings. These simpler objects, investigated extensively in the past \cite{Kak04apl,Gie07prb,Sch08prl,Lou09prl,Guo13prl,Lar14apl,Zho17prb}, can be naively viewed as constituent elements of the nanovolcanoes, see Fig.\,\ref{f1}(c). All nanovolcanoes exhibit a flat morphology and a slightly trapezoidal cross-sectional profile, as revealed by atomic force microscopy, see Fig.\,\ref{f1}(a).

The CPWs were prepared by e-beam lithography from a $55$\,nm-thick Au film sputtered onto a Si/SiO$_2$\,(200\,nm) substrate with a $5$\,nm-thick Cr buffer layer. The CPWs were covered with a $5$\,nm-thick TiO$_2$ layer, fabricated by e-beam lithography, for electrical insulation from the nanovolcanoes. The width and length of the active part of the CPWs were equal to $2D$ and $4D$ of the nanovolcano under study, respectively, see Fig.\,\ref{f1}(b). SWR measurements were taken in the frequency range $4$-$32$\,GHz with a bias magnetic field in the range $1.2$-$2$\,T, oriented perpendicular to the volcano plane. The field alignment error was less than $0.1^\circ$, which is too small to cause splitting of spin-wave modes in perpendicularly magnetized nanostructures \cite{Bun15nsr}. The high-frequency ac excitation was provided by a microwave generator and the transmitted signal was detected by a signal analyzer. The measurements were done with a bias magnetic field modulation amplitude of $1$\,mT and a frequency of $15$\,Hz, and a phase-sensitive recording of the microwave transmission.
\begin{figure}[t!]
    \centering
    \includegraphics[width=0.98\linewidth]{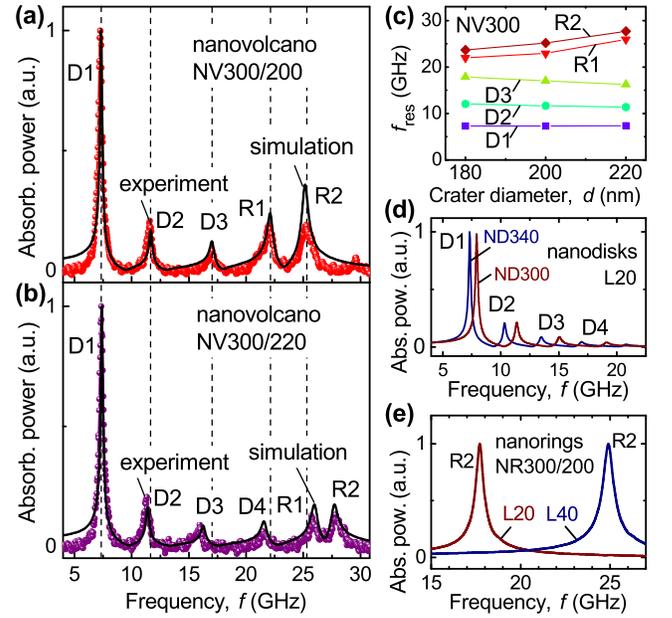}
    \caption{Measured (symbols) and calculated (lines) power absorption spectra for the nanovolcanoes (a) NV$300/200$ and (b) NV$300/220$.
    (c) Experimentally deduced resonance frequencies $f_\mathrm{res}$ vs crater diameter $d$ for the nanovolcanoes with $D=300$\,nm.
    Calculated power absorption spectra for
    (d) the nanodisks with thickness $L=20$\,nm and diameters $D = 300$\,nm and $340$\,nm, and
    (e) the nanorings NR$300/200$ with thicknesses $L=20$\,nm and $40$\,nm.
    In all panels, $H = 1.2$\,T.}
    \label{f2}
\end{figure}
\begin{SCfigure*}
    \centering
    \includegraphics[width=1.48\linewidth]{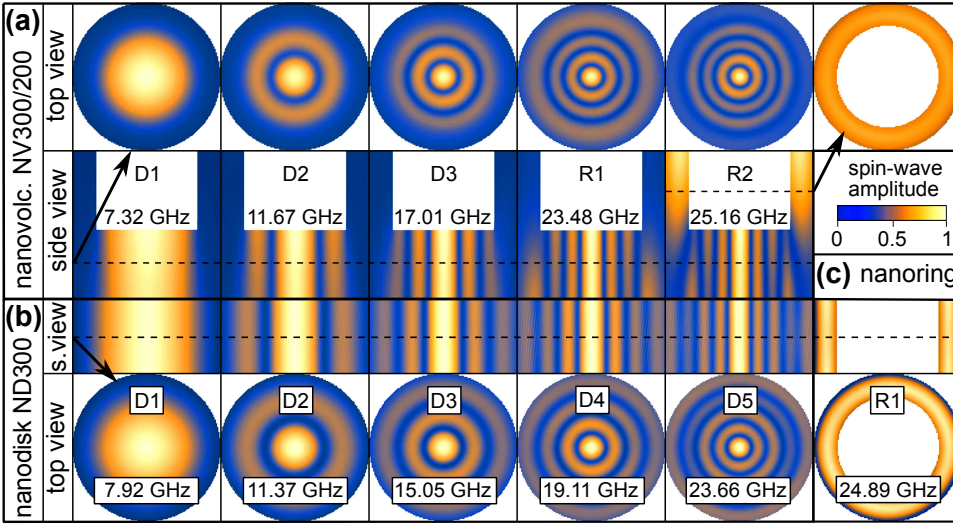}
    \caption{Calculated spatial dependences of the spin-wave eigen\-functions for
    (a) the $40$\,nm-thick nanovolcano NV$300/200$,
    (b) the $20$\,nm-thick nanodisk ND$300$, and
    (c) the $20$\,nm-thick nanoring NR$300/200$ at the out-of-plane bias magnetic field $1.2$\,T.
    The SWR modes correspond to those shown in Fig.\,\ref{f2}(a),
    (d), and (e).
    The mutual relation between the cross-sectional profiles are indicated by the dashed lines.}
    \label{f3}
\end{SCfigure*}


Figure \ref{f2}(a) and (b) present the experimentally measured microwave power absorption spectra for the nanovolcanoes NV$300/200$ and NV$300/220$ at $1.2$\,T. For both nanovolcanoes, the spectra contain the dominating low-frequency SWR modes (D1) at about $7.35$\,GHz and the pronounced higher-frequency SWR modes (R1 and R2). The increase of the crater diameter $d$ from $180$\,nm to $220$\,nm leads to a blue shift of the R1 and R2 peaks by about $4$\,GHz (see also Table \ref{t1}). Interestingly, these changes in the high-frequency part of the spectrum are accompanied by the opposite (though much weaker) shifts of the lower eigenfrequencies ($D_2$, $D_3$), and a nearly constant eigenfrequency $D_1$, as shown in Fig.\,\ref{f2}(c).

To identify the SWR modes associated with different parts of the nanovolcanoes, micromagnetic simulations were performed using the MuMax3 solver \cite{Van14aip}. For all nanovolcanoes the cell size was 2.5$\times$2.5$\times$2.5\,nm$^3$ and the damping parameter was $\alpha = 0.01$. The simulations were performed in two stages. Firstly, an equilibrium magnetic configuration of the system was reached by relaxing a random magnetic configuration for a given perpendicular bias field value. Subsequently, magnetization precession was excited by applying a small spatially uniform in-plane microwave field pulse. Finally, a fast Fourier transform was used to extract the normalized frequency spectra and the spatial dependences of the spin-wave eigenmodes. In the simulations, we used the saturation magnetization $M_\mathrm{s}$ and exchange stiffness $A$ values deduced for the disks with the same diameters and deposited with the same FEBID parameters \cite{Dob20nan} (Table \ref{t1}) and the assumed gyromagnetic ratio of $\gamma/2\pi = 3.05$\,MHz/Oe \cite{Tok15prl}. The decrease of $M_s$ and $A$ with the decrease of the disk diameter from $1\,\mu$m to $300$\,nm is associated with a decrease of the [Co+Fe] content from $80\pm3$\,at.\% to $63\pm3$\,at.\% and increase of the [C+O] content from $20\pm3$\,at.\% to $37\pm3$\,at.\% (as residues from the precursor in the FEBID process) \cite{Dob20nan}.

The calculated microwave absorption spectra for the nanovolcanoes NV$300/200$ and NV$300/220$ are shown by solid lines in Fig.\,\ref{f2}(a) and (b). The calculated spectra fit well with the experimentally measured ones. With increase of the diameter $d$, the R1 and R2 modes are shifted by a few GHz toward higher frequencies. The increase of $d$ has a very weak influence on the location of the SWR modes D1--D3. Observation of the enhancements afforded by the 3D extrusion in comparison to a pure nanodisk geometry is possible if one takes a naive view of a nanovolcano as a composite geometrical object composed of a $20$\,nm-thick nanodisk overlaid by a $20$\,nm-thick nanoring (geometry 1 in Fig.\,\ref{f1}(c)). The simulations predict the lowest-frequency mode for the nanodisk ND300 at a frequency $f_\mathrm{res}$ of $7.9$\,GHz (Fig.\,\ref{f2}(d)), which is about $600$\,MHz above the experimentally measured value for the equivalent diameter nanovolcano, NV300/200. In contrast, the calculated $f_\mathrm{res}$ of the 20\,nm-thick nanodisk ND340, which has a larger diameter of 340\,nm, matches well with the experimentally measured $f_\mathrm{res}=7.32$\,GHz of the NV300/200 nanovolcano. This is to say that the lowest-frequency eigenmode of a model nanovolcano, i.e. a  $20$\,nm-thick disk (volcano basement) overlaid by a $20$\,nm-thick ring, can be obtained in simulations for a $20$\,nm-thick disk which has a \emph{larger effective diameter} $D_\mathrm{eff}=340$\,nm. That is, extension of a 2D nanomagnet into the third dimension allows one to engineer the lowest eigenfrequency of a nanodisk by using a 3D nanovolcano which has an about 30\% \emph{smaller footprint}. This phenomenon can be explained by the complex internal field $H_\mathrm{int}$ distribution in the 3D nanovolcanoes (Fig.\,\ref{f1}(d)), which is responsible for the dynamic demagnetization tensor resulting in a decrease of the effective spin pinning. Thus, while the spin precession angle in thin disks is zero at the disk edges \cite{Kak04apl}, the spin precession angle is non-zero in the case of nanovolcanoes.

We have also calculated $f_\mathrm{res}$ for $20$\,nm- and $40$\,nm-thick nanorings NR$300/200$ at $1.2$\,T (Fig.\,\ref{f2}(e)). For the $20$\,nm-thick nanoring NR$300/200$ (geometry 1 in Fig.\,\ref{f1}(c)) $f_\mathrm{res} = 17.75$\,GHz is very far away from the modes R1 ($22.02$\,GHz) and R2 ($25.15$\,GHz) of the equivalently sized nanovolcano NV$300/200$. This is because of the essentially larger internal field in the volcano's ring area as compared to the $20$\,nm-thick ring (Fig. \ref{f1}(d)). By contrast, for a $40$\,nm-thick nanoring NR$300/200$\,nm at $1.2$\,T (geometry 2 in Fig.\,\ref{f1}(c)) $f_\mathrm{res} = 24.9$\,GHz is very close to the measured R2 eigenfrequency ($25.1$\,GHz) of the equivalently sized nanovolcano NV$300/200$\,nm. In this way, the R2 peak can be ascribed to the ring mode of a nanoring with the same thickness as the nanovolcano and the width equal to that of the ring encircling the nanovolcano crater (geometry 2 in Fig. \ref{f1}(c)).

The calculated spatial dependences of the spin-wave eigenmodes for the nanovolcano NV300/200 (Fig.\,\ref{f2}(b)) are shown in Fig.\,\ref{f3}(a). At first glance, the eigenmodes in the nanovolcano resemble spin-wave ``drum modes'' known for nanodisks saturated in the perpendicular geometry \cite{Kak04apl,Dob20nan}. These ``drum modes'' are approximately described by Bessel functions of the zeroth order \cite{Kak04apl} and are shown in Fig.\,\ref{f3}(b) for the $20$\,nm-thick nanodisk ND300 for comparison. Such interpretation explains why the volcano low-frequency modes only weakly depend on the crater diameter, see Fig. \ref{f2}(c). Indeed, a closer look at the D1-D3 mode profiles reveals that spin waves in the nanovolcano are confined under the volcano crater which acts as a ``concentrator'' for spin waves. The role of the ring overlying the volcano basement disk becomes more decisive for higher-frequency eigenmodes. Specifically, the R1 mode has the maximum amplitude at the bottom surface of the nanovolcano while the R2 mode has the maximal amplitude in the upper part of the ring around the crater. This elucidates why the location of the R1 and R2 modes is very sensitive to the width $(D-d)/2$ of the ring encircling the volcano crater. The exchange interaction for the rings with widths $\simeq50$\,nm essentially contributes to the R2 frequency (the exchange length \cite{Gus05prb} $\lambda = \sqrt{2A/(\mu_0 M_\mathrm{s}^2)}$ in Co-Fe-FEBID is about $5$\,nm). Accordingly, the localization of the R2 mode in the ring is not a result of the reduced dipolar pinning \cite{Wan19prl}, but is rather a result of the interplay of the exchange and internal magnetic fields.

\begin{figure}[t!]
    \centering
    \includegraphics[width=0.98\linewidth]{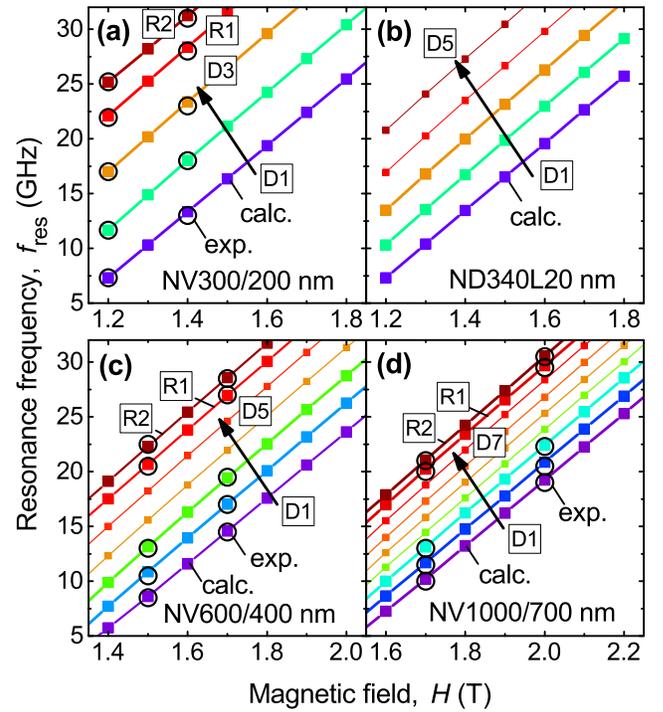}
    \caption{Resonance frequencies versus bias magnetic field as deduced from SWR measurements (circles) and calculated numerically (squares) for (a) the nanovolcano NV$300/220$\,nm, (b) nanodisk ND$340$L20\,nm, (c) nanovolcano NV$600/400$\,nm, (d) and nanovolcano NV$1000/700$\,nm. The dominating modes are indicated with larger symbols and thicker lines. The slope of straight lines is determined by $\gamma/2\pi = 3.05$\,MHz/Oe.}
    \label{f4}
\end{figure}

Finally, with increase of the nanovolcano diameter $D$, the number of the SWR modes per given frequency interval increases, see Fig.\,\ref{f4}. This can be understood as a reduction of the system sizes leads to a stronger confinement of spin waves and the associated larger mode separation in the magnon frequency spectrum. With increase of the magnetic field, the spectra are shifted towards higher frequencies as a whole, without qualitative changes in the structure of the spectra and the relative positions of the peaks. This can be understood on the basis of the nearly linear dependence $f_\mathrm{res}(H)\simeq (\gamma / 2\pi)H$ for axially symmetric nanoelements magnetized along the symmetry axis \cite{Kak04apl}. The slope of all straight lines $f_\mathrm{res}(H)$ in Fig.\,\ref{f4} is the same and it is determined by the gyromagnetic ratio $\gamma / 2\pi$.

To summarize, we have introduced nanovolcanoes as on-demand engineered nano-architectures for 3D magnetism and magnon spintronics. The $40$\,nm-thick Co-Fe nanovolcanoes with diameters down to $300$\,nm were fabricated by the direct-write FEBID technique and studied by perpendicular SWR spectroscopy. The spin-wave eigenfrequencies of the nanovolcanoes have been demonstrated to notably differ from the eigenfrequencies of the nanodisks and nanorings they are built from, because of the strongly non-uniform internal magnetic field. The experimental findings were elucidated with the aid of micromagnetic simulations which indicate that the rings encircling the volcano craters lead to an effective confinement of the lower-frequency eigenmodes under the volcano crater while the highest-frequency eigenmodes are confined in the ring area. Accordingly, extension of 2D nanodisks into the third dimension allows one to engineer their lowest eigenfrequencies by using 3D nanovolcanoes having about 30\% smaller footprints. By varying the crater diameter by $\pm20$\,nm, we have demonstrated frequency tuning of about $\pm2$\,GHz for the ring modes without affecting the lowest spin-wave eigenfrequency. The presented nanovolcanoes can be viewed as multi-mode resonators with potential applications in telecom-frequency filters. In addition, the engineered spin-wave frequency spectra make them prospective platforms for 3D magnonics and inverse-design magnonic devices.\\

\footnotesize{OVD and SLC acknowledge the Austrian Science Fund (FWF) for support through Grant No. I 4889 (CurviMag).
The Portuguese team acknowledges the Network of Extreme Conditions Laboratories-NECL and Portuguese Foundation of Science and Technology (FCT) support through Project Nos. NORTE-01-0145-FEDER-022096, POCI-0145-FEDER-030085 (NOVAMAG), and EXPL/IF/00541/2015.
NZ and AVC acknowledge the Austrian Science Fund (FWF) for support through Grant No. I 4917.
KG acknowledges support by IKERBASQUE (the Basque Foundation for Science). The work of KG was supported by the Spanish Ministerio de Ciencia, Innovacion y Universidades grant FIS2016-78591-C3-3-R.
Support through the Frankfurt Center of Electron Microscopy (FCEM) is gratefully acknowledged.
Further, support by the European Cooperation in Science and Technology via COST Action CA16218 (NANOCOHYBRI) is acknowledged.\\[2mm]
The data that supports the findings of this study are available within the article.
}


%

\end{document}